\documentclass{ws-ijmpd}
\usepackage[super,compress]{cite}
\usepackage{graphicx}
\usepackage{epstopdf}
\usepackage{amsfonts}
\usepackage{amssymb}
\usepackage{epsfig}
\usepackage{hyperref}

\usepackage{dcolumn}
\usepackage{bm}
\usepackage{amsmath}
\usepackage{epstopdf}
\usepackage{amsfonts}
\usepackage{amssymb}
\usepackage{epsfig}
\usepackage{tabularx}
\usepackage{color}
\usepackage{hyperref}

\begin{document}


%
\catchline{}{}{}{}{}
%

\title{Acoustic modes of pulsating axion stars: Non-radial oscillations}

\author{Grigoris Panotopoulos}

\address{Centro de Astrof{\'i}sica e Gravita{\c c}{\~a}o-CENTRA, Departamento de F\'isica, 
\\
Instituto Superior T\'ecnico-IST,
Universidade de Lisboa-UL,
\\
Av. Rovisco Pais 1, 1049-001 Lisboa, Portugal
\\
\href{mailto:grigorios.panotopoulos@tecnico.ulisboa.pt}{\nolinkurl{grigorios.panotopoulos@tecnico.ulisboa.pt}} 
}

\author{Il{\'i}dio Lopes}

\address{Centro de Astrof{\'i}sica e Gravita{\c c}{\~a}o-CENTRA, Departamento de F\'isica, 
\\
Instituto Superior T\'ecnico-IST,
Universidade de Lisboa-UL,
\\
Av. Rovisco Pais 1, 1049-001 Lisboa, Portugal
\\
\href{mailto:ilidio.lopes@tecnico.ulisboa.pt}{\nolinkurl{ilidio.lopes@tecnico.ulisboa.pt}} 
}

\maketitle

\begin{history}
\received{Day Month Year}
\revised{Day Month Year}
\end{history}

\begin{abstract}
We compute the lowest frequency non-radial oscillation modes of dilute axion stars. The effective potential that enters into the Schr{\"o}dinger-like equation, several associated eigenfunctions, and the large as well as the small frequency separations are shown as well.
\end{abstract}


\ccode{}

\section{Introduction}

Modern observational data coming from different sides of Astrophysics and Cosmology show that dark matter (DM) \cite{zwicky,rubin} dominates the non-relativistic matter in the Universe \cite{turner}. The DM problem comprises one of the biggest challenges in modern theoretical Cosmology, and although several potential candidates exist \cite{taoso}, the origin and nature of DM still remains a mystery. For reviews on DM and on its detection searches see e.g. \cite{DM1,DM2,DM3,DM4,DM5}.

The collisionless DM paradigm based on weakly interacting massive particles \cite{kolb&turner} works very well on large (cosmological) scales ($\ge Mpc$), but encounters several problems at low (galactic) scales, like the core-cusp problem, the diversity problem, the missing satellites problem and the too-big-to-fail problem \cite{Tulin}. Self-interacting DM was introduced long time ago \cite{steinhardt} in an effort to resolve, or at least to alleviate, the small scale crisis of collisionless DM.

The QCD axion \cite{axion1,axion2,Marsh} is a pseudo-Goldstone boson that solves the strong CP problem of Quantum Chromodynamics (QCD) in an elegant way, while at the same time it is an excellent DM candidate \cite{taoso}. Indeed, the most compelling solution to the strong CP problem, which can be stated as "why is the $\Theta$ parameter in QCD so small?", is the solution proposed by Peccei and Quinn (PQ) \cite{PQ1,PQ2}. According to the PQ mechanism an additional global, chiral symmetry is introduced, now known as the PQ symmetry, which is spontaneously broken at the PQ scale $F_\alpha$, and it suffers from a chiral anomaly. It is a well-established fact from Quantum Field Theory that in this case there must be a pseudo-Goldstone boson associated with this symmetry and also characterized by a small non-vanishing mass. The pseudo-Goldstone boson associated with the PQ scale is the axion yet to be detected. If the axion exists it has the potential to resolve simultaneously the strong CP problem of QCD and the DM problem in Cosmology. 

Boson stars are are gravitationally bounded collections of bosons. Boson stars have been studied in \cite{BS1,BS2,BS3,BS4,BS5,BS6,BS7}, see also \cite{chavanis1,chavanis2} for Newtonian self-gravitating Bose-Einstein condensates. The maximum mass for bosons stars in non-interacting systems was found in \cite{BS1,BS2}, while in \cite{BS3,BS4} it was pointed out that self-interactions can cause significant changes. In \cite{BS5,BS6} the authors constrained the boson star parameter space using data from galaxy and galaxy cluster sizes. Regarding axion stars in particular, the possibility of the formation of axion stars was considered for the first time in \cite{tkachev}, and since then axion stars have been studied in \cite{chavanis3,Visinelli,Braaten}.

It is well-known that the properties of stars, such mass and radius, depend crucially on the equation-of-state (EoS), which unfortunately is poorly known. This has changed, however, during the last years thanks to recent advances in Helioseismology and Asteroseismology in general, that are becoming precise science. Studying the oscillations of stars and computing the frequency modes, at least in principle, offer us the
opportunity to probe the interior of the stars and learn more about the EoS, since the precise values
of the frequency modes are very sensitive to the thermodynamics and the internal structure and composition of the star \cite{Turck}.

A stable sphere of gravitationally bounded axions is usually known as an axion star. These stars belong to one of two types depending of their energy density profile: a diluted axion star in which the equilibrium results from the balance between the kinetic pressure of the axions and gravitational attractive force; and a dense axion star in which the balance is due to the equilibrium between axion self-interaction and the gravitational force \cite{Visinelli,Braaten}. The radius of a diluted axion star can be several orders of magnitude lower than the radius of a dense axion star with the same stellar mass. For instance, an axion star with a mass of $10^{-17}$~$M_\odot$ has a radius of $10^{-10}$~$R_\odot$ or  $10^{-2}$~$R_\odot$ in the case of a diluted or a dense axion star, respectively, with $R_\odot$ being the solar radius and $M_\odot$ being the solar mass. In this study, our fiducial stable axion star has an approximate mass of  $10^{-14}$ $M_\odot$ and radius of $10^{-6}$  $R_\odot$ \cite{chavanis3}.

An axion star, like a normal star, is a self-gravitating sphere which will experiment non-radial oscillations when perturbed by some internal or external mechanism. In some astrophysical scenarios, the axion star can occasionally become unstable, in which by gravitational collapse the star finds a new stable configuration \cite{Katz}. This gravitational contraction can be accompanied by non-radial oscillations. Among the various astrophysical scenarios that can possibly explain the origin of non-radial oscillations, the merger of a boson star with a compact star, such as a black-hole or a neutron star, are among the most compelling examples. In the latter case the time to coalescence increases due to tidal effects between the stars, therefore, in that way increasing the time during which the axion star will experiment non-radial oscillations by tidal effects. As a matter of fact, some of the observational phenomena found in pulsars, x-ray binaries, magnetars and binary neutron star systems could also be explained by assuming that the axion star is one of the two (or even both) components of the stellar system. This new line of research is largely motivated by the recent discoveries in gravitational wave astronomy. 

In recent 3D numerical relativity simulations that focused towards understanding and finding the leading  astrophysical signatures of a neutron star with a axion star merger~\cite{Day}, the authors found that the axion star goes through a series of gravitational perturbations, during  which the tidal forces of the neutron star could excite non-radial oscillations of the axion star. This work established that such merger event could be observed by a variety of observational channels: gravitational waves, optical and near-infrared electromagnetic waves, radio flares, fast radio bursts, gamma ray bursts, and neutrino emissions. As pointed out by \cite{Cardoso}, using only gravitational wave data it will be difficult to distinguish between the mergers of two compact stars (such as two neutron stars), and the mergers of boson stars (such as axion stars). The same holds for the electromagnetic spectrum of such mergers. Therefore, the detailed study of non-radial oscillations could help astronomers to better understand the underlying physics in these physical processes.

Nowadays, only a restricted number of astrophysical signatures have been 
proposed to detect axion stars. It is usually assumed that axions stars can be located in galactic or stellar environments with a strong magnetic field, such as the ones found in the proximity of a neutron star. These magnetic fields have been observed in many types of stars, as well as in all types of galaxies and galaxy clusters, Milky Way included. In such astrophysical scenarios, axions from the axion star will interact with the local magnetic field producing a continuum flux of electromagnetic radiation or a monochromatic wave. The electromagnetic radiation is produced as a result of the coupling of an axion with two virtual photons which convert the axion to real photons in the magnetic background \cite{Primakoff,Wilczek}. There are two cases for the production of axion signals, worth mentioning in this work: 

-- (i)  the  monochromatic radio wave signal with a frequency 
proportional to the mass of the axion particle. In \cite{Bai1} it was found that the conversion rate of axions in photons is sufficiently large to produce a strong signal able to be detected by existing radio telescopes and by the future SKA radio telescope. This mechanism has been proposed to explain some of the observed fast radio bursts.

-- (ii) the continuum axion electromagnetic spectrum~\cite{Bai2}. In this case, the axions generated in the interior of the star (normal or otherwise) are converted almost directly to the x-ray radiation as the axions pass through the atmosphere of the star via the interaction of the axion with the magnetic field of the star.  In \cite{Zioutas} it was found that in the case of the Sun, the x-ray spectrum is almost a blackbody radiation with a maximum at the energy of 3 keV and a mean energy value of 4.2 keV.

These examples mentioned above give us a glimpse of future possibilities to look for non-radial oscillation signals coming from axion stars.

In this work we develop a similar strategy to study the internal properties of these hypothetical axions stars, which are on the list of a future astronomical observational programs \cite{Kelley,Bai3}. For previous works on radial oscillations of stars see e.g. \cite{Cox,chanmugan1,Frandsen, Aerts,Hekker,chanmugan2,kokkotas,hybrid,basic,LP} and references therein.

Our goal in the present article is to study for the first time 
non-radial oscillations of axion stars. Our work is organized as follows: after this introduction, we present the EoS and the solution of hydrostatic equilibrium in section two, and the equation for non radial small adiabatic acoustic perturbations formulated as a quantum mechanical problem in the third section. Our numerical results are presented in section four, and finally we conclude in the last section. We work in natural units in which $c=\hbar=1$. In these units all dimensionfull quantities are measured in GeV, and we make use of the conversion rules $1 m = 5.068 \times 10^{15} GeV^{-1}$, $1 kg = 5.610 \times 10^{26} GeV$ and $1 sec = 1.519 \times 10^{24} GeV^{-1}$ \cite{guth}.

\section{Hydrostatic equilibrium}

\subsection{Equation-of-state}

The axion self-interaction potential is generated by non-perturbative instanton effects given by \cite{PQ1}
\begin{equation}
V(\phi) = \Lambda^4 \left[ 1 \pm cos\left( \frac{\phi}{F_\alpha} \right) \right].
\end{equation}
We see that it is a periodic potential with height $\Lambda$ and width $F_\alpha$, while the axion mass
$m_\alpha = \Lambda^2/F_\alpha$ determines the curvature in the vicinity of the extrema of the potential.
The Peccei-Quinn scale is bounded both from below and from above from cosmological and astrophysical considerations, $(10^8-10^9)~GeV \leq F_\alpha \leq 10^{12}~GeV$ \cite{kim1,kim2}. The QCD scale, however, is experimentally known quite accurately \cite{QCDscale1,QCDscale2}, and its world average for 3 quark flavours is given by \cite{prd}
\begin{equation}
\Lambda^{(3)} = (332 \pm 17)~MeV.
\end{equation}

The axion self-interactions can be read off expanding the axion potential in powers of $\phi$
\begin{equation}
V_{int}(\phi) \simeq \frac{\Lambda^4 \phi^4}{24 F_\alpha^4} - \frac{\Lambda^4 \phi^6}{720 F_\alpha^6}
\end{equation}
In a cold dilute boson gas almost all particles are in the ground state described by the Gross-Pitaevskii equation \cite{BEC1,BEC2,BEC3}, also known as non-linear Schr{\"o}dinger equation, and the collection of axions is described by a polytropic EoS $P(\epsilon)=K \epsilon^2$ with polytropic index $n=1$, or $\gamma=2$, while the constant $K$ is given entirely in terms of the QCD scale \cite{chavanis3}
\begin{equation}
K = \left( \frac{1}{2 \Lambda}  \right)^4.
\end{equation}

\begin{table}
\caption{Axion star's lowest frequencies $\nu_{l,n}$  of non-radial oscillations, for
dipole ($l=1$), quadrupole ($l=2$) and octupole ($l=3$)  modes  (for QCD scale $\Lambda=330~MeV$).}	
\begin{tabular}{l | c c c}	
{\footnotesize overtones}  & $\nu_{1,n}$ & $\nu_{2,n}$  & $\nu_{3,n}$  \\
$n$ & $(mHz)$ & $(mHz)$ & $(mHz)$ \\
		\hline
		\hline
		0  & 0.8441 & 0.8457&1.3274 \\
		1  & 1.7409 & 1.7426 &2.2288  \\
		2  & 2.5397& 2.5456 &3.0588  \\
		3  & 3.3096 &  3.3155 & 3.8555 \\
		4  & 4.0657 & 4.0676&4.6332  \\
		5  & 4.8140 &4.8110 &5.3990 \\
		6  & 5.5576 & 5.5579 &6.1568  \\
		7  & 6.2980 & 6.3032 &6.9090 \\
		8  & 7.0362 &7.0359 &7.6571 \\
		9  & 7.7727 &7.7774 &8.4021 \\
		10 & 8.5081 & 8.5076 &9.1448 \\
		11 & $\cdots$ & 9.5719 &9.8855 \ \\
		12 &   &10.3079 &10.6248  \\
		13 &   & 11.0432&11.3628  \\
		14 &   &11.7777  &12.0999 \\
		15 &   &$\cdots$ &12.8361 \\
		16 &   & &13.5716 \\
		17 &   & &14.3066 \\
		18 &   & &$\cdots$ \\
	\end{tabular}
	\centering
	\label{table:FreqSet}
\end{table}

\subsection{Structure equations}

Since the axion star is non-relativistic described by a polytropic EoS, to study the hydrostatic equilibrium one has to solve the non-relativistic version of the Tolman-Oppenheimer-Volkoff (TOV) equations \cite{tolman,OV}
\begin{equation}
m'(r) = 4 \pi r^2 \epsilon(r)
\end{equation}
for the mass function, and
\begin{equation}
P'(r) = - \epsilon(r) \frac{m(r)}{r^2}
\end{equation}
for the pressure, where the prime denotes differentiation with respect to the radial coordinate $r$.
Combining these two equations we can derive a single second order differential equation, known as the
Lane-Emden equation \cite{textbook}
\begin{equation}
\frac{d}{dx} \left(x^2 \frac{d \theta}{dx} \right) = -x^2 \theta
\end{equation}
with the initial conditions
$ \theta(0)  =  1  $ and $
{d \theta}/{dx}(0)  =  0$,
where the new variables are defined as follows
\begin{eqnarray}
x & = & \frac{r}{a} 
\end{eqnarray}
and
\begin{eqnarray}
\theta & = & \frac{\epsilon}{\epsilon_c}
\end{eqnarray}
with $\epsilon_c$ being the central energy density, while $a$ is given by
$a = \sqrt{K/2 \pi}$.
It is easy to verify that the solution
\begin{equation}
\theta(x) = \frac{sin(x)}{x}
\end{equation}
satisfies both the Lane-Emden equation and the initial conditions.
Therefore, the energy density as a function of the radial coordinate is given by
\begin{equation}
\epsilon(r) = \epsilon_c \frac{sin(r/a)}{(r/a)}.
\end{equation}
The above equation is valid for the radius varying from $r=0$ until the first zero of the function $\epsilon(r) $, therefore the function $\epsilon(r) $ varies between $\epsilon_c$ and 0. Finally, the mass $M$ and the radius $R$ of the star are given by
\begin{eqnarray}
M & = & 4 \pi \epsilon_c a^3 \int_0^\pi dx x^2 \theta(x) \\
R & = & \pi a
\end{eqnarray}
Clearly, only the mass of the star depends on the central energy density, while the radius is fixed. This happens only in the special case $n=1$, whereas in general both $M$ and $R$ depend on the central energy density. This can also be seen in the mass-to-radius profile for Newtonian boson stars with repulsive forces as shown 
in the work of Chavanis and collaborators (see Fig.~2 of \cite{chavanis1} and Fig.~4 of \cite{chavanis2}).

\begin{figure}[ht!]
	\centering{\includegraphics[scale=0.75]{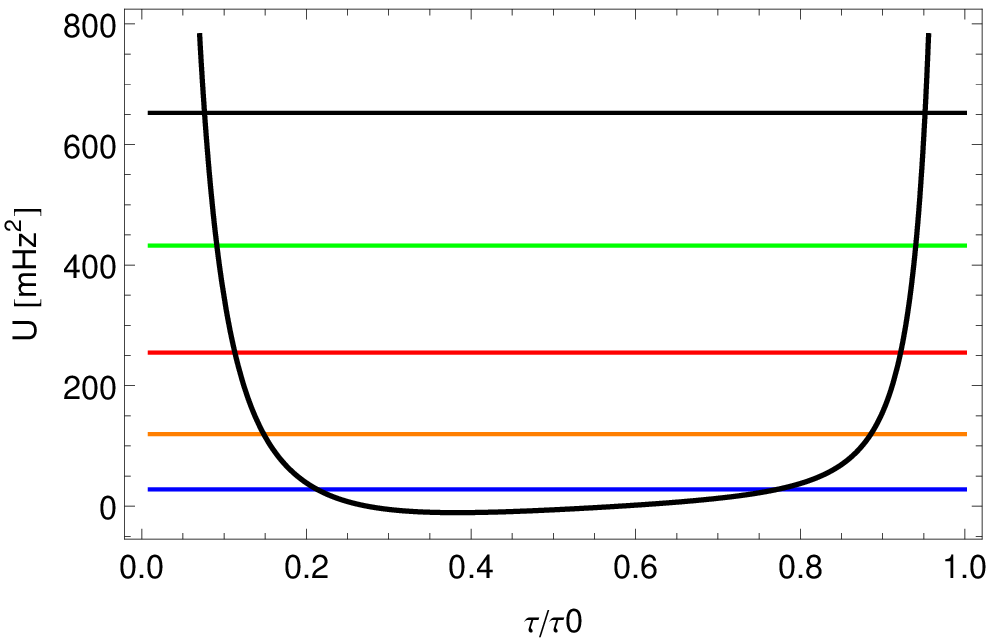}}
	\centering{\includegraphics[scale=0.75]{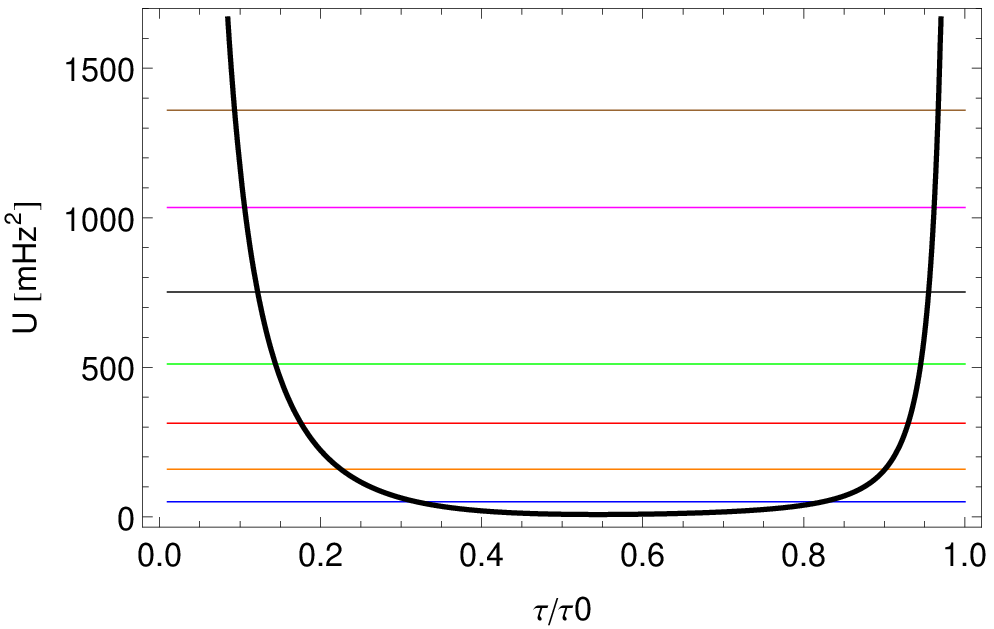}}
	\centering{\includegraphics[scale=0.75]{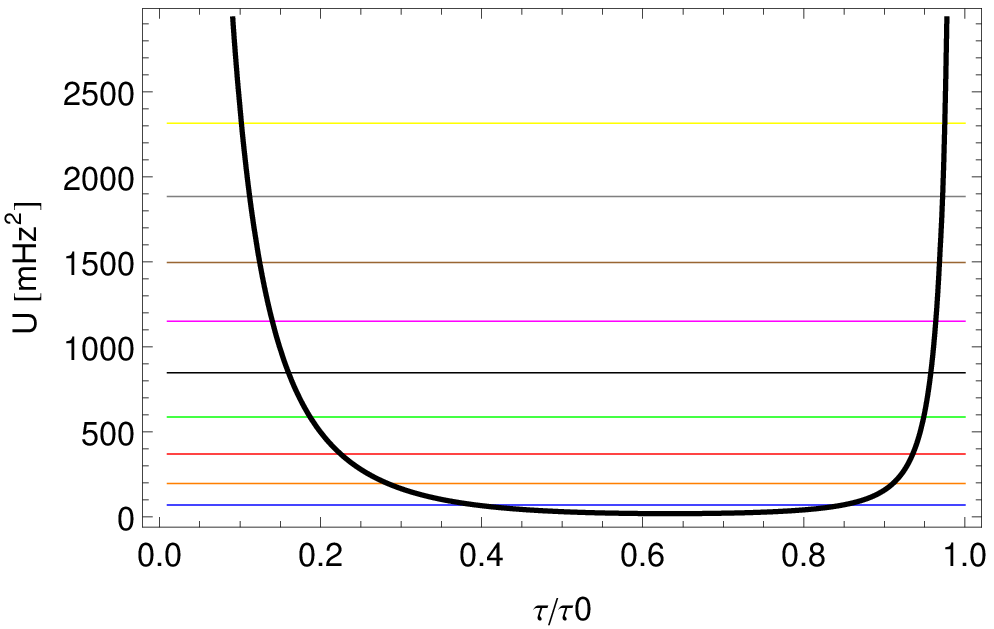}}
	\caption{Effective potential $U_l$ (in $(mHz)^2$) versus dimensionless time $\tau/\tau_0$, and the first oscillation 
		frequency modes for $\Lambda=330~MeV$ and $l=1,2,3$. The horizontal line give the frequencies of lowest 
		overtones of each $l$ mode set: {\bf (a)} {\it top panel}, dipole modes ($l=1$) with n=0,1,2,3,4;
		{\bf (b)} {\it middle panel}, quadrupole modes ($l=2$) with n=0,1,2,3,4,5,6;
		{\bf (c)} {\it bottom panel}, octupole modes ($l=3$) with n=0,1,2,3,4,5,6,7,8.}
	\label{fig:potentialU} 
\end{figure}

\begin{figure}[ht!]
	\centering{\includegraphics[scale=0.65]{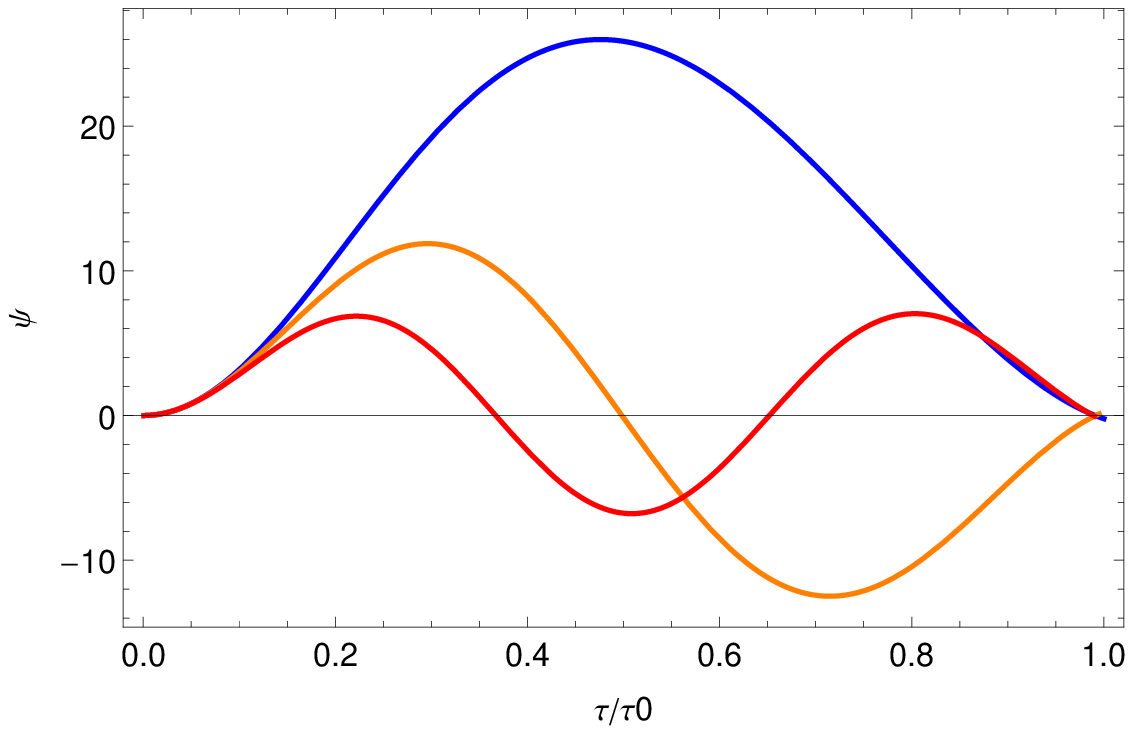}}
	\centering{\includegraphics[scale=0.65]{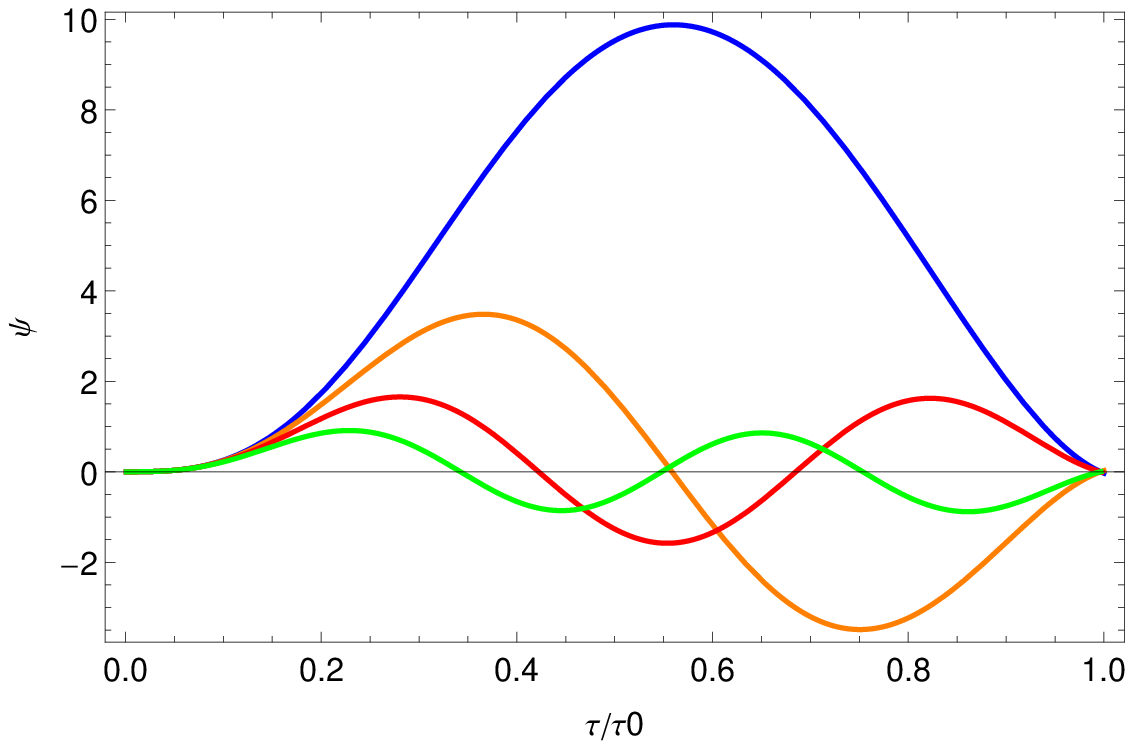}}
	\centering{\includegraphics[scale=0.65]{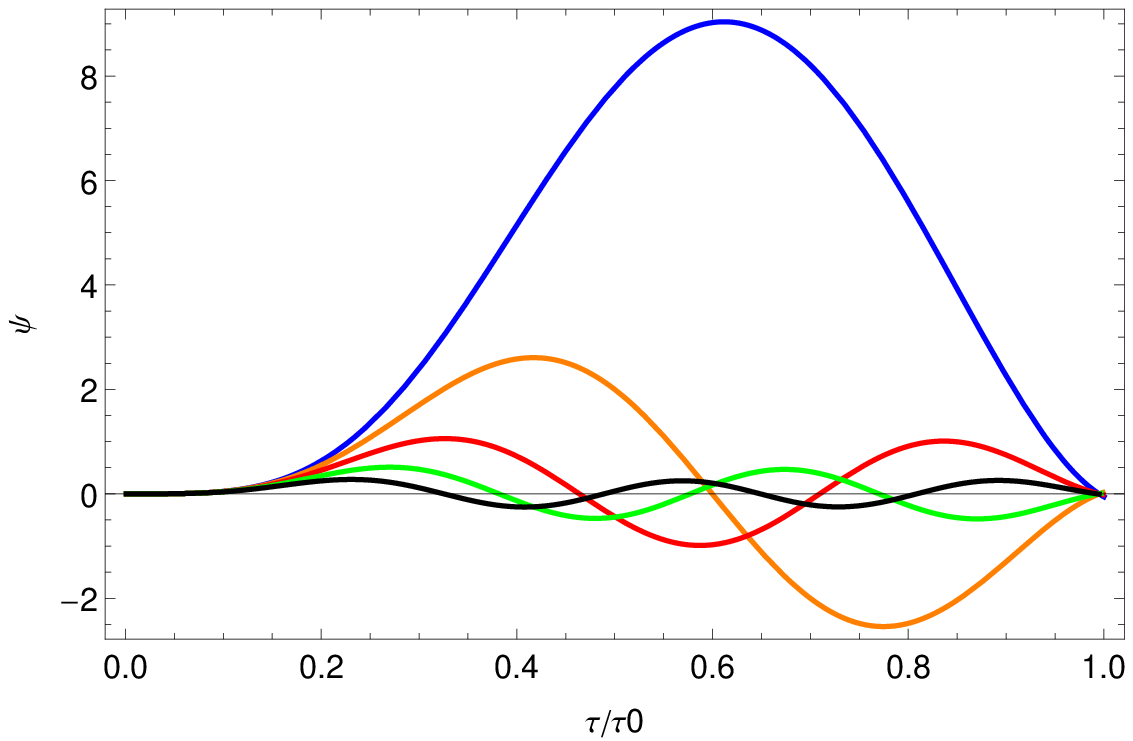}}
	\caption{The eigenfunctions $\psi_{l,n}$  versus dimensionless time $\tau/\tau_0$:
		{\bf (a)} {\it top panel}, $\psi_{1,n}$ for the fundamental (n=0, blue), the first (n=1, orange) and second (n=2, red) excited 
		dipole ($l=1$) modes; 
		{\bf (b)} {\it middle panel},  $\psi_{2,n}$  for the fundamental mode (n=0,blue), the first (n=1,orange), second (n=2,red) and 
		third (n=3,green) excited  quadrupole  ($l=2$) modes.
		{\bf (c)} {\it bottom panel}, $\psi_{3,n}$ (n=0,1,2,3,4) for the fundamental mode (n=0,blue), the first  (n=1,orange), second (n=2,red), third (n=3,green) and fourth (n=4, black) excited octupole ($l=3$) modes.
		The frequencies are shown on table~\ref{table:FreqSet}.}
	\label{fig:eingfunctions} 	
\end{figure}

\section{Non-radial oscillations}

\subsection{Equations for the perturbations}

Linear adiabatic acoustic perturbations in the Cowling approximation \cite{cowling}, where the perturbations of the gravitational potential are neglected, are described by the following equation \cite{ledoux}
\begin{equation}
\zeta''(r) + \left( \frac{2}{r} + \frac{2 \epsilon'(r)}{\epsilon(r)} \right) \zeta'(r) + \left(\frac{\omega_{n,l}^2}{c_s^2} - \frac{l (l+1)}{r^2} \right)  \zeta(r) = 0
\end{equation}
where 
 $c_s$ is the speed of sound defined by $c_s^2 = dP/d\epsilon$, $\omega_{n,l}$ ($=2\pi\nu_{n,l}$) are the discrete eigenvalues, $l > 0$ is the degree of angular momentum (or degree of the mode), and $n=0,1,2,...$ is the overtone number (or radial mode).

The Sturm-Liouville boundary value problem at hand can be treated equivalently as a quantum mechanical problem by recasting the second order differential equation for $\zeta$ into a Schr{\"o}dinger-like equation \cite{Lopes1,Lopes2,Lopes3} of the form
\begin{equation}
\frac{d^2 \psi}{d \tau^2} + \left[ \omega^2 - U_l(\tau) \right] \psi = 0.
\label{eq:Psi}
\end{equation}
Introducing the functions
\begin{equation}
A(r) = \frac{2}{r} + \frac{2 \epsilon'(r)}{\epsilon(r)},
\end{equation}
\begin{equation}
H_l(r) = c_s(r)^2 \frac{l (l+1)}{r^2},
\end{equation}
and
\begin{equation}
P(r) = A(r) c_s(r) - c_s'(r).
\end{equation}
The new variables $\tau$ and $\psi$ are defined as follows
\begin{equation}
\psi(r)  = \frac{\zeta(r)}{u(r)}
\end{equation}
where $u$ satisfies the condition $u'/u= -P/(2 c_s)$, and $\tau$ is the acoustic time
\begin{equation}
\tau  =  \int_0^r c_s^{-1}(z) dz.
\end{equation}
Finally, the effective potential is found to be
\begin{equation}
U_l(r) = H_l(r) + \left( \frac{P(r)}{2} \right)^2+\frac{c_s(r) P'(r)}{2},
\label{eq:Ul}
\end{equation}
and we thus obtain the effective potential as a function of the acoustic time in parametric form $\tau(r), U_l(r)$.

The expression for the acoustic potential (equation~\ref{eq:Ul}) above defines the inner and outer turning points where the acoustic wave propagation changes from oscillatory behaviour to exponential decay. For a given acoustic wave with frequency $\omega$ there are two radii ($r_{in}$ and $r_{out}$, such $r_{in}< r_{out}$ ), for which $ U_l (r)=\omega^2$ (see equation~\ref{eq:Psi}). These radii are known as inner ($r_{in}$)  and outer ($r_{in}$) turning points. An acoustic wave with a given frequency will oscillate in the regions of the stellar interior where $\omega^2\ge U_l( r) $ and will be evanescent in the regions in which $\omega^2\le U_l( r) $ (cf. Figures~\ref{fig:potentialU}a,b,c). This analysis illustrates how the formation of standing waves depends of the detailed structure of the outer layers of the star, namely, the density, the sound speed, $\cdots$, and the first and second derivatives of these quantities. For that reason, even if the axion star does not have a well-defined surface, it still has the  ability to form standing waves~\cite{Liebling}.

\subsection{Numerical results}

We consider for the QCD scale the value $\Lambda=330~MeV$ \cite{QCDscale1,QCDscale2}, and a central pressure that corresponds to an axion star with a mass $M=1.86 \times 10^{-14}~M_{\odot}$ and radius $R=6.9~km$. Therefore, the order of magnitude for the fundamental mode is expected to be \cite{textbook} $\omega_0 \propto \sqrt{M/R^3} = 2.74~mHz$. In the existing literature it has been pointed out by several authors that very light dark matter clumps may be detected by Pulsar Timing Arrays\cite{PTA1,PTA2,PTA3}. Furthermore, space-based detectors, such as LISA, will probe gravitational waves (GWs) in the mHz regime \cite{lisa}, which is precisely the case investigated in this work. Since the axion stars considered here are very light, the amplitude of the GW from an individual object is expected to be extremely low. However, if axion stars in our galaxy pulsate collectively in huge numbers at the same time, the expected power of the stochastic gravitational wave background may be significantly enhanced.

The spacing in frequencies $\nu_{n,l}=\omega_{n,l}/(2 \pi)$ for the highly excited modes is computed by \cite{Lopes4,ilidio}
\begin{equation}
\nu_0 = \left[ 2 \int_0^R \frac{dr}{c_s(r)} \right]^{-1},
\end{equation}
where $c_s$ is the sound speed given by $c_s^2 = dP/d \epsilon$.
Since the background solution is known we obtain for the spacing the numerical value $\nu_0 = 0.73~mHz$. What is more, since we have an exact analytical solution for the hydrostatic equilibrium we obtain the following expression
\begin{equation}
\nu_0 = \left( \frac{\pi}{2K^3} \right)^{1/4} \frac{\sqrt{M}}{a_0} \label{basic}
\end{equation}
where $a_o$ is a numerical constant, such that  $a_o=\int_0^\pi \sqrt{z/sin{z}}\,  dz \approx 5.8993$. We notice that the constant spacing $\nu_0 $ is proportional to $\Lambda^3$, and it scales with the square root of the mass of the star. For the mass assumed here we obtain the same numerical value, $\nu_0 = 0.73~mHz$, obtained before.

We computed the lowest oscillation mode frequencies summarized in  table~\ref{table:FreqSet} for $l=1,2,3$ respectively. We see that the fundamental mode is of the expected order of magnitude. Furthermore, although the numerical values of the individual frequencies increase with $l$, the constant spacing at higher excited modes is always the same, given by \ref{basic}, irrespectively of the value of the angular degree, as observed in Fig.~\ref{fig:spectrum} below. Introducing $\tau_0=\tau(R)=11.4~min$, the effective potential $U_l$ that enters into the Shr{\"o}dinger-like equation versus dimensionless $\tau/\tau_0$ is shown in Figures~\ref{fig:potentialU}a,b,c for $l=1,2,3$ respectively, while the eigenfunctions or several modes are shown in Figures~\ref{fig:eingfunctions}a,b,c. 

For an acoustic wave with frequency $\omega$ propagating inside the star, its eigenfunction will oscillate in the region  $ r_{in} < r < r_{out}$ (cf. the effective potential in Figures~\ref{fig:potentialU}), and it will decay exponentially in the remaining regions, as shown in the Figures~\ref{fig:eingfunctions}a,b,c. The amplitude the eigenfunctions at the surface of the star results from the combined action of two physical processes:  the mechanism responsible for the excitation of the non-radial modes inside the star, and the extension of the evanescent region near the stellar surface. Actually, even if all the acoustic non-radial oscillations were excited by a mechanism that is independent of the frequency and degree of the mode, these oscillations will have quite distinct amplitudes at the surface. Indeed, the amplitude of the eigenfunctions at the surface depends of the ability of the mode crossing throughout the evanescent region near the surface (equivalent to the tunnel effect in quantum mechanics), and emerge at the stellar surface with minor decay. In this work, we found that acoustic modes with larger frequency and large degree have a smaller "near the surface" evanescent region, consequently leading to eigenfunctions with larger amplitudes at the  surface.

We recall that in a Sturm-Liouville boundary value problem the number of zeros of the eigenfunctions corresponds to the overtone number $n$, namely the first excited mode corresponds to $n=1$ and has only one zero, the second excited mode corresponds to $n=2$ and has two zeros, while the fundamental mode does not have zeros and corresponds to $n=0$. In Figure~\ref{fig:eingfunctions}a we show the eigenfunctions  $\psi$ for $l=1$ corresponding to the fundamental mode, $n=0$ (blue), and the first two excited modes, $n=1$ (orange), and $n=2$ (red). In Fig.~\ref{fig:eingfunctions}b,c we show the associated eigenfunctions for $l=2$ and $3$ respectively for the fundamental and several excited modes. Fig.\ref{fig:Norma}   shows in logarithmic scale a comparison between the shape of the effective potential with different $l$, i. e., $U_1$, $U_2$ and $U_3$.

\begin{figure}[ht!]
\centering{\includegraphics[scale=0.75]{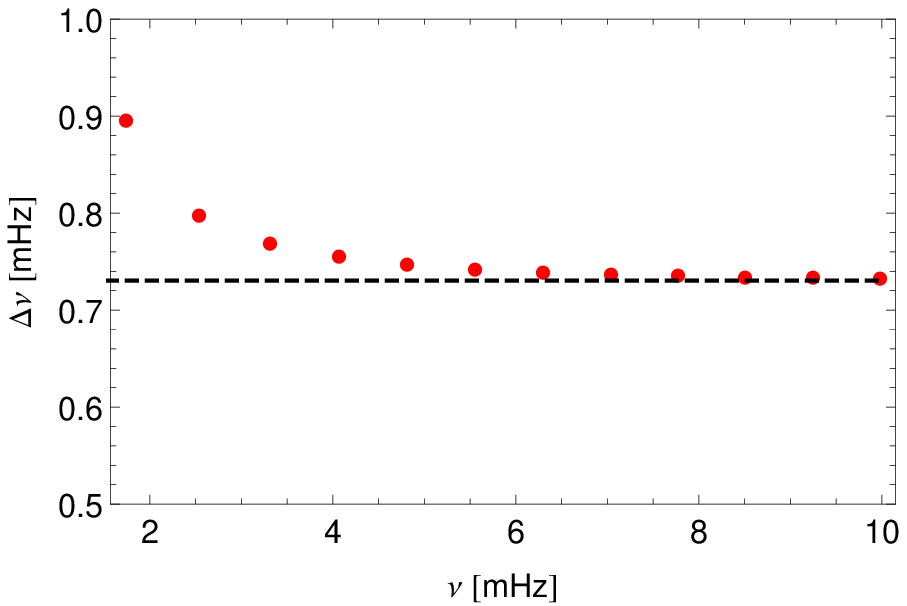}}
\centering{\includegraphics[scale=0.65]{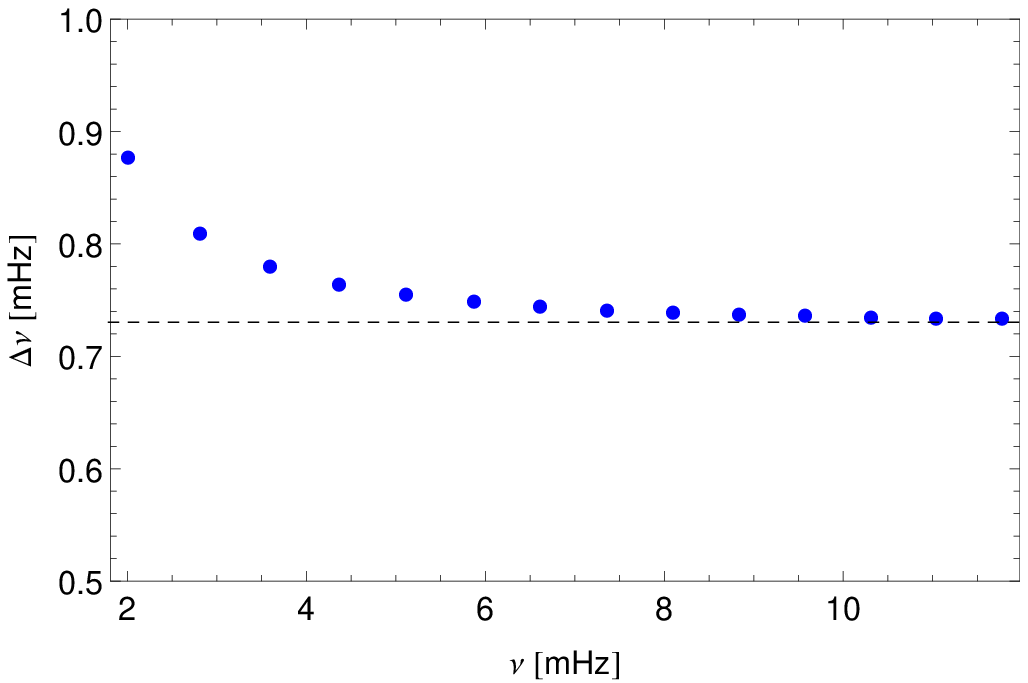}}
\centering{\includegraphics[scale=0.65]{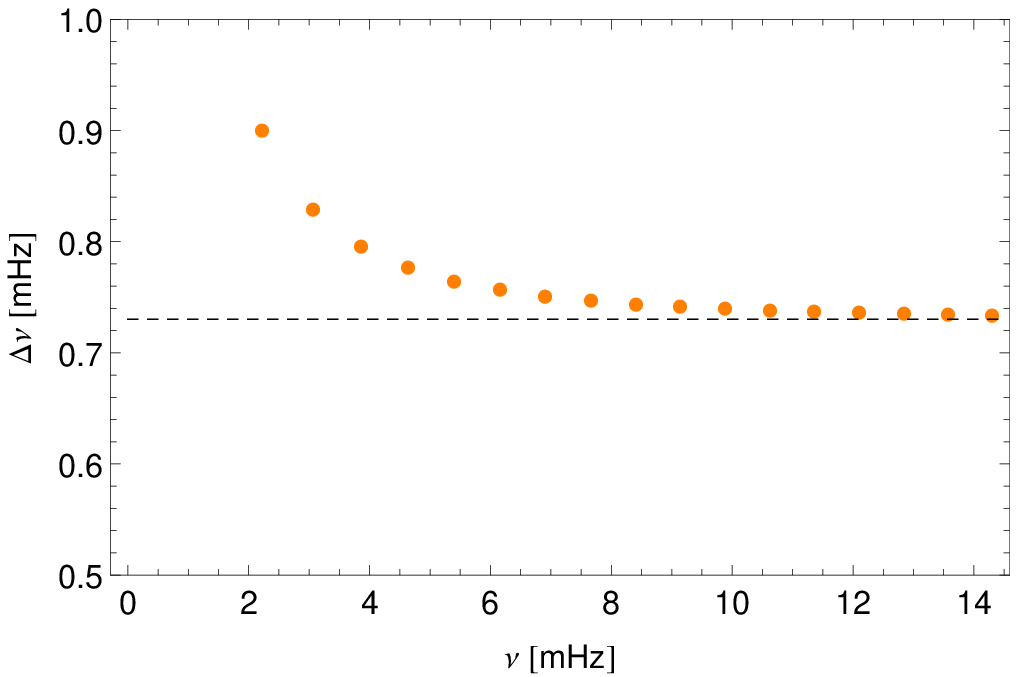}}
\caption{$\Delta \nu_{l,n}$ versus $\nu_{l,n}$ (both in $mHz$) for the lowest acoustic modes: 
{\bf (a)} {\it top panel},  dipole ($l=1$) modes; 
{\bf (b)} {\it middle panel},  quadrupole  ($l=2$) modes.
{\bf (c)} {\it bottom panel}, octupole ($l=3$) modes.
The dotted line corresponds to $\nu_o$.
The frequencies are shown on table~\ref{table:FreqSet}.}
\label{fig:spectrum} 	
\end{figure} 
 
 \begin{figure}[ht!]
 	\centering
 	\includegraphics[scale=0.8]{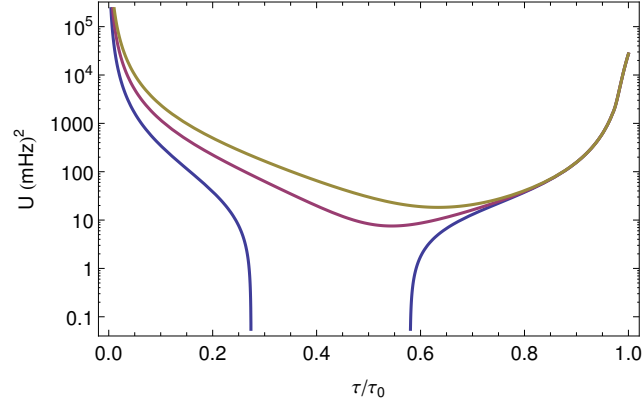}
 	\caption{Effective potential (in $(mHz)^2$) for $l=1,2,3$ versus dimensionless time $\tau/\tau_0$
 		for $\Lambda=330~MeV$.}
 	\label{fig:Norma} 	
 \end{figure}

\begin{figure}[ht!]
	\centering
	\includegraphics[scale=0.85]{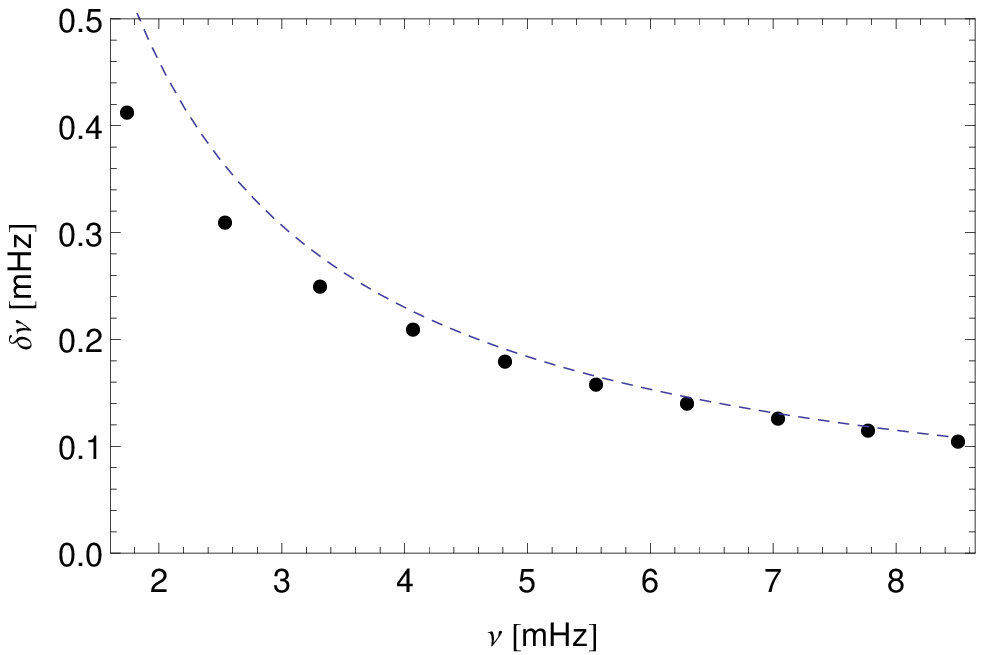}
	\caption{Small frequency separation $\delta \nu_{n,l}$ versus $\nu_{n,l}$ (both in $mHz$) for the lowest acoustic modes for $\Lambda=330~MeV$ and $l=1$ and $l=3$.}
	\label{fig:small} 	
\end{figure}

Next we compute the large frequency separation for a given $l$
\begin{equation}
\Delta \nu_{l,n} = \nu_{l,n+1}-\nu_{l,n},
\end{equation}
shown in Figures~\ref{fig:spectrum}a,b,c for $l=1,2,3$ respectively, and for $\Lambda=330~MeV$. At higher order $m$ the spectrum exhibits a series of equally spaced modes, with the spacing being $0.73 mHz$, precisely as expected. As $l$ increases we need to compute even higher modes in order to reach the spacing $\nu_0$. We can also compute the small frequency separation relating modes corresponding to two different values of $l$ as follows \cite{book}
\begin{equation}
\delta \nu_{l,n} = \nu_{l,n}-\nu_{l+2,n-1}
\label{eq:deltanu}
\end{equation}
shown in Fig.~\ref{fig:small}. The dashed line corresponds to the theoretically expected small frequency separation valid for higher excited modes given by \cite{book}
\begin{equation}
\delta \nu_{l,n} \simeq - (6+4l) \frac{\nu_0}{4 \pi^2 \nu_{l,n}} \: \int_0^R dr \frac{c_s'(r)}{r}
\end{equation}
for $l=1$ and $\nu_0 = 0.73 mHz$, while the integral is computed to be
\begin{equation}
\int_0^R dr \frac{c_s'(r)}{r} = -4.97~mHz
\end{equation}
A simple analysis of figure~\ref{fig:Norma} shows that the  difference  $\delta \nu_{l,n}$ (equation~\ref{eq:deltanu}) of two close frequency modes (for instance with  $l=1$ and $l=3$) is sensitive to the core of the star. 

Before concluding our work a final comment is in order.
Although here we have assumed an axion star mass $M \sim 10^{-14}~M_{\odot}$, by increasing the central energy density we can increase the mass by several orders of magnitude to reach the Hz frequency range, while the radius remains the same as already explained in subsection 2.2. Moreover, it is easy to verify that the object remains dilute and non-relativistic. This leads to qualitatively similar results, the only difference being the higher values of the frequencies, as expected from the simple formula quoted before in the text, $\omega_0 = \sqrt{M/R^3}$. In that case the passage of the axion star next to a compact object, which may help to excite the non-radial oscillations by tidal forces, requires a $\sim sec$ long observable signal to detect its Hz modulation, which is of the same order of magnitude of the passage time scale.

\section{Conclusions}

We have studied non-radial oscillations of axion stars. The axion, a pseudo-Goldston boson with a small finite mass, has the potential to solve the strong-CP problem of QCD in an elegant way, and at the same time the DM problem of Cosmology. For three different values of the degree of angular momentum $l=1,2,3$ we have computed the lowest frequencies, several associated eigenfunctions, and the effective potential in the equivalent description in terms of a Schr{\"o}dinder-like equation. Both the large and the small frequency separations are shown as well. In all three cases, like in the radial oscillation case, for the higher excited modes the large separation tends to a constant determined entirely by the QCD scale.


\section*{Acknowlegements}

We are grateful to the anonymous reviewers for valuable comments and suggestions. It is a pleasure to thank K.~Clough, C.~Moore and D.~Hilditch for enlightening discussions. The authors thank the Funda\c c\~ao para a Ci\^encia e Tecnologia (FCT), Portugal, for the financial support to the Center for Astrophysics and Gravitation-CENTRA, Instituto Superior T\'ecnico, Universidade de Lisboa, through the Grant No. UID/FIS/00099/2013.


\end{document}